\documentclass[11pt]{article}

\usepackage{natbib,graphicx,color}
\usepackage{float}
\usepackage{graphicx}
\usepackage{colortbl}
\usepackage{geometry}

\begin{document}

\title{Modeling Complex Interactions in a Disrupted Environment:
          Relational Events in the WTC Response\thanks{This work was supported by NSF awards CMMI-2027475 and SES-1826589.}}
\author{Scott Leo Renshaw\thanks{Department of Sociology, University of California, Irvine} and Selena M. Livas\thanks{Department of Sociology, University of California, Irvine} and Miruna G. Petrescu-Prahova\thanks{School of Public Health, University of Washington} and Carter T. Butts\thanks{Departments of Sociology, Statistics, Computer Science, and EECS, University of California, Irvine}\thanks{To whom correspondence should be addressed; \texttt{buttsc@uci.edu}}
}     

\date{September 2021}
\maketitle
\newtheorem{lemma}{Lemma}[section]

\label{firstpage}

\maketitle

\begin{abstract}
When subjected to a sudden, unanticipated threat, human groups characteristically self-organize to identify the threat, determine potential responses, and act to reduce its impact.  Central to this process is the challenge of coordinating information sharing and response activity within a disrupted environment.  In this paper, we consider coordination in the context of responses to the 2001 World Trade Center disaster.  Using records of communications among 17 organizational units, we examine the mechanisms driving communication dynamics, with an emphasis on the emergence of coordinating roles.  We employ relational event models (REMs) to identify the mechanisms shaping communications in each unit, finding a consistent pattern of behavior across units with very different characteristics.  Using a simulation-based ``knock-out'' study, we also probe the importance of different mechanisms for hub formation.  Our results suggest that, while preferential attachment and pre-disaster role structure generally contribute to the emergence of hub structure, temporally local conversational norms play a much larger role.  We discuss broader implications for the role of microdynamics in driving macroscopic outcomes, and for the emergence of coordination in other settings.
\end{abstract}

\section{Introduction}

It is a common ``disaster myth'' that, when faced with a sudden threat, social groups lacking specific preparation will either passively wait for rescue, or fly into a state of uncontrolled panic \citep{tierney.et.al:aaapss:2006}.  In fact, it is far more common for those in harm's way to take immediate action to assess the threat, determine appropriate response measures, and take the initiative to carry them out \citep{auf.der.heide:ch:2004}.  Such complex behavior under adverse and disrupted circumstances poses significant problems of \emph{coordination}: to accomplish it, groups must compile information regarding the evolving situation (as well as relevant background knowledge), identify actions that need to be taken (and the resources needed for those actions), and direct behavior to ensure that requisite actions are performed with a minimum of task interference.  Central to successful coordination is communication, the structure of which can either facilitate or inhibit performance (as has been known at least since the pioneering studies of Bavelas and colleagues \citep[e.g.][]{bavelas.barrett:per:1951}).  At a more microdynamic level, effective communication also depends upon shared interaction norms, whether embedded in formal protocols or in conventional cultural practices, without which information passing becomes extremely difficult.

While this emergence of coordination in response to threat is well-established, the exact mechanisms by which it is accomplished \emph{in situ}, and the relative importance of those mechanisms for successful response in practice, remain the subject of inquiry.  Among the challenges in studying this phenomenon have been the relative paucity of detailed data on communication dynamics in disrupted settings, and until recently a lack of principled modeling strategies for inferring the driving mechanisms behind interaction processes from observational data (particularly in naturalistic settings wherein many different types of mechanisms may be simultaneously at play).  Progress within the Relational Event Modeling (REM) paradigm \citep{butts:sm:2008} has greatly lowered this latter barrier, making it practical to investigate complex microdynamics within a statistically principled framework.  There remains, however, the empirical challenge of identifying and analyzing cases of emergent coordination in response to external threat.

In this paper, we contribute to this latter goal via an investigation of coordination in response to the 2001 World Trade Center disaster.  Building on the work of \citet{butts:sm:2008}, who studied six small communication networks from this event, we here analyze 17 relational event systems from both specialist and non-specialist responders.  Considering a range of candidate mechanisms, we identify those present in each network and estimate their effects. Further, we employ simulation-based analysis of the estimated models to probe the relative importance of different mechanisms for the emergence of hub structure - a critical coordinative adaptation in these groups.  As we show, beneath the diversity of these responding groups lies considerable consistency, with the vast majority of communication mechanisms operating in similar ways across networks when present (though not all mechanisms operate in all networks).  While we find that well-known mechanisms such as preferential attachment and the prominence of pre-disaster coordinative roles do consistently contribute to the formation of hub structure, a much larger fraction in fact emerges from the action of temporally local communication norms, which have the side effect of creating conversational ``inertia'' that leads to large differences in communication activity.  The macro-level structure of the WTC communication networks is thus seen to arise in large part from microdynamic mechanisms.

The remainder of the paper proceeds as follows. First, we give a brief overview of communication in disaster context, along with the use of interpersonal radio devices (on which the present case is based) as a medium for communication. Next, we provide some details on the differences between specialist and non-specialist networks and how network size may influence communication, followed by a discussion of the mechanisms potentially involved in hub formation. We then describe our dataset and the methods used for model selection, analysis, and the simulation knock-out experiment. This is followed by a presentation of the results for our relational event models, as well as our simulation study. Finally, we end the paper with a discussion of the implications of our findings for communication and coordination in disrupted settings more generally, and for future studies.  Our paper is also accompanied by an R package \citep{rteam:sw:2021} with the complete WTC radio data set, thereby facilitating further analysis of this rich historical case.

\section*{Background}

Responses to disasters depend upon a complex interplay of formal (i.e., institutional) and informal factors \citep{quarantelli:ch:1966,dynes:bk:1970,stallings:ch:1978}. This interplay is perhaps nowhere more evident than in the domain of responder communication, where technical constraints, formal roles, and standard operating procedures must cope with responders' shifting demands for information and the capacity to provide it. An important aspect of the total response process is the emergence of communication networks that promote information transmission and coordination. The structural characteristics of such networks, as well as the mechanisms that facilitate their formation, are of crucial importance to researchers and practitioners alike.

By definition, disasters are exceptional events, in which conventional social processes are subject to substantial disruption.  While losses are perhaps the most salient characteristic of disasters, Drabek's (\citeyear{drabek:bk:1986}) classic synthesis of findings on disaster response emphasizes the ``accidental or uncontrollable'' nature of disaster events (p. 7), and the extent to which they exceed the capacity of conventional mechanisms for managing disruption. Indeed, a central feature of the US Federal Emergency Management Agency's definition of disaster is the requirement that the event ``cannot be managed through the routine procedures and resources of government'' \citep{fema:tr:1984}. Uncertainty and disruption of routines are therefore an important aspect of the social responses to disaster events. Organizationally, such disruptions of routine generate a high-uncertainty environment in which coordination demands escalate, while infrastructure (both human and technical) degrades. The problem of ``many people trying to do quickly what they do not ordinarily do, in an environment with which they are not familiar'' \citep[p77]{tierney:par:1985} generates the potential for confusion and task interference, particularly where tasks are time-critical and resources are limited. Negotiating such difficulties, acquiring information about losses and ongoing hazards, and other coordinative tasks require a high degree of interpersonal communication. Thus, actors responding to a disaster depend on the emergence (or retention, if pre-existing) of communication networks, which can convey information from those who have it to those who need it without placing excessive demand on the communicants \citep{drabek:bk:1986}.

Also central to the nature of disaster communication is the time frame in which the communication takes place. Disaster researchers conventionally divide the ``life cycle'' of a disaster into several periods or phases \citep{fischer:bk:1998}, distinguished by characteristic patterns of activities and events, e.g. a ``pre-impact period'' before the hazard manifests; an ``impact period'' during which the hazard is active; a ``response period'' in which damage is contained and survivors are attended to; and a ``recovery period'' in which attempts are made to restore conditions to a stable state.  Although communication is critical in all periods, the impact and response periods constitute a crucial interval in which survivors attempt to respond to ongoing hazards, search and rescue operations begin, and emergency response organizations respond to the scene and attempt to coordinate their efforts.  Within that interval, it is common in practice to refer to an ``emergency phase'' in which immediate, time-sensitive action is required to react to an active hazard.  Communication in this period plays a vital role in facilitating situational awareness, and in coordinating response activities. At the same time, such communication is made more difficult by disruption of conventional resources, roles, and routines, the high opportunity cost of engaging in communication versus task performance, and the high cognitive load facing communicants in what is typically a confusing, fast-changing, and possibly threatening environment.  Understanding the emergent dynamics of emergency phase communication thus has the potential to shed light on how groups organize in response to threats within a high-pressure setting that differs greatly from everyday conditions.  To date, the lack of detailed data from real emergencies has been a major barrier to such understanding, and the methodology needed to make use of such data has only become available in recent years. In this paper, we capitalize on the unique assets of the World Trade Center dataset (described below). The 2001 World Trade Center (WTC) disaster stands as one of the largest communication-coordination ``emergency phase'' related events in recent history. With this data and the increasingly widely-used relational event framework, we intend to provide novel insights into communications during the emergency phase of disaster response.

\subsection*{Interpersonal Radio Communications}

From the mid-20th century onward, radio communication via portable devices has been a critical tool for coordination among responders \citep{mcelroy:qst:2005}, despite numerous limitations \citep{auf.der.heide:bk:1989}. In the immediate post-impact period, when communications infrastructure may be degraded and alternate systems have not yet been deployed, hand-held radio devices serve to connect responders in the field to one another. In addition, the relatively low cost of hand-held transceivers in the modern context makes this technology accessible to organizations which do not specialize in emergency response activities. Since the first responders to any disaster are those who happen to be at the impact site, interpersonal radio communication is an important ``workhorse'' tool for improving coordination in the emergency phase of a disaster. This raises the question, however, of how responders -- especially those who are not specialized in emergency response -- actually use radio communication during the emergency phase of a disaster.  At the most basic label, radio communications require the use of fairly rigid communication protocols to avoid confusion due to cross-talk; these are typically codified into a set of practices or standard operating procedures (SOP) that involve systematic identification of the sending party and intended receiving party, formal acknowledgments of contact and receipt, etc.  In prior work on a subset of the WTC radio networks, \citet{butts:sm:2008} found strong evidence for the prominence of radio SOP via pronounced and systematic participation shift effects.  Organizations using radio communications may also delegate coordination tasks to specially designated individuals (e.g. dispatchers), potentially leading to a more centralized structure in which individuals largely interact with the institutionally designated coordinator rather than directly with each other.  We discuss this further below.

\subsection*{Specialized vs. Non-Specialized Responders}

It is commonly posited that behavioral responses to disasters by ``ordinary people'' within the initial phase of a hazard event are deviant and chaotic, by contrast with the disciplined and efficient actions of emergency response organizations \citep{tierney.et.al:aaapss:2006}.  In practice, however, organizational responses are not necessarily better coordinated than those of others on the scene \citep{fischer:bk:1998}, and the latter may indeed prove quite effective \citep{auf.der.heide:ch:2004}.  As Fischer (\citeyear{fischer:bk:1998}) famously observed, variation in extent of prior planning, rehearsal of plans, and previous experience with similar events plays a large role in determining which organizational responses are successful and which fall short (a point also noted by e.g. \citet{drabek:bk:1986,auf.der.heide:bk:1989}). That said, since the first individuals and organizations to respond to a disaster are those who happen to be present when the impact occurs, it is not necessarily the case that the true ``first responders'' will be trained for or equipped to deal with the event at hand. Much emergency-phase response activity is thus improvised, but it does not follow that such activity will be disorganized or ineffective \citep{wachtendorf:diss:2002,mendonca.et.al:jccm:2014}. Repurposing of existing communication networks, together with the emergence of new ones, may result in highly structured interaction patterns. The question of how such networks develop, then, and their dependence upon pre-disaster organization, is of clear importance to understanding responder communication in the immediate post-impact period.  

In the case of radio communications by teams of WTC responders, \citet{butts.et.al:jms:2007} distinguish between groups of ``specialist'' responders - police, security, or other personnel who are specifically trained and organized to respond to emergencies (if not disasters) - and ``non-specialist'' responders (e.g., maintenance personnel) who are present and active at the scene, but whose conventional organizational structure and practices are not intended for emergency response. Specialist and non-specialist groups responding to a disaster share goals such as getting individuals to safety \citep{quarantelli:sp:1960, mileti.et.al:bk:1975, abe:me:1976, noji:ch:1997}, but specialist groups may also be charged with other objectives (e.g., securing the area, taking actions to prevent or mitigate emerging or ongoing hazards, coordinating with other organizational units, etc.) that may pose different communicative or coordinative challenges. \citet{butts.et.al:jms:2007} found that the time-aggregated networks of communication among specialist and non-specialist groups share structural features, with relatively minor quantitative differences but a high level of overall similarity. In a dynamic analysis of the six smallest WTC networks, \citet{butts:sm:2008} also found broad similarity in the communication patterns of specialist and non-specialist responders. However, some consistent differences have also been identified; for instance, an analysis of robustness of the WTC networks to attack \citep{fitzhugh.butts:sn:2021} found that specialists were consistently more dependent upon individuals in institutionalized coordinative roles to maintain connectivity (and hence more vulnerable to their removal).  This last suggests a potential difference between specialist and non-specialist networks in their hub organization, a matter that we revisit in our simulation study below.

\subsection*{Emergent Coordination and Institutionalized Coordinative Roles}

In cross-sectional analyses of the WTC data, \citet{petrescu-prahova.butts:ijmed:2008} show evidence that both specialist and non-specialist networks are held together by a relatively small number of highly central actors (\emph{coordinators}).  While some of these actors occupy \emph{institutionalized coordinative roles} (e.g., manager, dispatcher) that formalize their status as a coordinator under nominal conditions, the relationship between such formal roles and actual coordination is imperfect.  Although individuals occupying an institutionalized coordinator role (ICR) are more likely to become coordinators than those without such roles, the majority of coordinators were found to be \emph{emergent} (i.e., to lack an ICR).  This raises the question of what drives the emergence of coordination during the unfolding emergency, and the relationship of those drivers to ICR status.  Using REMs, we can directly probe the effect of ICR occupancy on communication behavior patterns, as well as differences in these behaviors between specialist and non-specialist networks (an effect not examined in previous work).   Using simulation, we can further analyze the impact of ICR effects on the formation of communication hubs at the network level, allowing us to investigate the extent to which behavioral differences involving ICRs do or do not ramify into differences in structural position (and to which this varies between specialist and non-specialist networks).  

\subsection*{Network Size}

Another factor that may plausibly impact WTC radio communications is network size.  The 17 networks vary over a large range, from 24 to 256 individuals (mean 127), reflecting substantial differences in the complexity of events within the group of actors and the cognitive load entailed in keeping track of events within the group.  This may lead to differences in the relevant mechanisms driving communication between networks; for instance, egocentric ``tracking'' of distinct communication threads beyond the currently active conversation is likely to be of substantially greater importance in large networks with many different lines of activity, while adherence to local conversational norms may be greater in smaller groups with fewer risks of interruption.  Likewise, some mechanisms may have greater impact in networks of larger or smaller size.  For instance, preferential attachment may have a greater impact in larger networks, as there may be less \emph{a priori} clarity in who is available to communicate in a large group and a correspondingly greater signal value of visibility.  For ICRs, the task of coordinating activity becomes more difficult in larger networks, as they must coordinate a larger number of actors as well as relay information to more individuals, while presumably receiving information from more sources as well.  This may plausibly reduce the gap between ICRs and other members of the network, as emergent coordinators step in to perform tasks that ICRs cannot.  Below, we examine the question of whether these and other mechanisms operate differently in networks of different size.

\section*{Potential Hub-Forming Mechanisms}

A consistent finding in prior descriptive analyses of the WTC networks \citep{butts.et.al:jms:2007,petrescu-prahova.butts:ijmed:2008,fitzhugh.butts:sn:2021} is that specialist and non-specialist networks alike are held together by a relatively small number of individuals occupying hub-like coordinator roles.  These positions are distinguished by high degree, betweenness, and total communication volume, and are reflective of individuals who played outsized roles in coordinating activity during the WTC event.  As noted above, the majority of these positions are \emph{emergent,} in that that do not correspond to ICR membership, although those in ICRs are disproportionately likely to occupy coordinator/hub roles.  This raises the question of whence these hubs come: what are the dynamic mechanisms that drive the hub-focused organization of the WTC networks, and are these mechanisms consistent across organizational units?  Here, we consider three categories of mechanisms that could plausibly account - individually, or in tandem - for hub formation in the WTC radio networks.

\subsection*{Preferential Attachment}

One of the best known mechanisms driving the emergence of hubs in social networks is preferential attachment \citep{price:jasis:1976}.  Early empirical studies of preferential attachment go back as far as the 1950s and 1960s \citep{merton:s:1968,simon:b:1955}, with the emphasis on ``cumulative advantage'' processes in settings such as citation networks, where new entrants tend to cite papers already cited by others; the derivation of power law degree distributions from such processes is due to \citet{price:jasis:1976}, with much later rediscoveries several decades later by researchers outside the social network community.  In the context of a relational event process with fixed vertices, preferential attachment can be understood as an increasing propensity to direct events towards vertices with a larger share of prior communication (i.e., larger total communication volume), as implemented e.g. by \citet{butts:sm:2008,gibson.et.al:po:2019,gibson.et.al:sn:2021}. In a radio communication setting, those who speak first are known to all persons attending to the channel to be active, thus making them a likely target for incoming communications. This may in turn lead to their receiving more calls, their responding and thus getting more air time. Over time, this process may produce a positive feedback loop in which those who are active early on (possibly for idiosyncratic reasons) end up becoming hubs. In the case of the WTC disaster, there are reasons why we might expect this pattern to appear. First, communication is occurring through the use of radio, meaning the first responders cannot see who they are communicating to, or identify who is available to communicate with. The simplest way to know who is available to respond is by noting who has already spoken, making the likelihood of their response high. Secondly, in the early stages of the disaster, individuals are cognitively overloaded and often disoriented; the cognitive load can be eased by choosing to communicate to those who are known to be already talking rather than trying to call individuals who may not be able to respond.

The social mechanism of preferential attachment may also be more likely to drive the formation of hubs in non-specialist networks when compared to their specialist counterparts. Specialist networks have protocols in place that often dictate a pre-planned chain of command for whom to contact in the case of a disaster, which may make them less likely to experience preferential attachment. Non-specialists lack these pre-existing norms and training, leading to a greater reliance on the ``call whom you hear'' heuristic that drives preferential attachment.

\subsection*{Conversational Inertia}

A second mechanism that could account for the emergence of coordination hubs is that of \emph{conversational inertia}, which arises from the turn-taking structure of conversational norms (in this case, radio SOP).  Parties engaged in conversation tend to remain in conversation, either because of the back-and-forth of call/response dynamics, or because of other normative actions such as sequential address (in which the same party continues to be the sender), baton-passing (in which the receiving party becomes the sender), ``piling on'' (in which the current receiving party becomes the subject of additional incoming communications), etc. tend to keep one or both currently communicating parties involved.  This produces a form of autocorrelation that can potentially amplify other sources of variation (including random events) in communication volume: individuals who, for whatever reason, are drawn into conversation tend to stay there, accumulating substantially more communication volume than those who are not.  While not normally thought of as a driver of hub formation, it is apparent that such micro-level processes could have this effect, and given the central importance of radio SOP to effective communication in events such as the WTC disaster we investigate its potential impact.  As we shall show, this impact turns out to be substantially greater than might be supposed.

\subsection*{Institutionalized Coordinator Roles (ICR)}

The last hub forming mechanism we investigate here is differential behavior by and towards individuals occupying institutionalized coordinative roles.  As discussed above, persons in ICRs (whom we simply refer to as ``ICRs'' where there is no danger of confusion) are expected, to receiver and relay information, direct action, and perform other duties requiring high levels of communication.  ICRs have also been found to be more likely to occupy coordinator roles in practice \citep{petrescu-prahova.butts:ijmed:2008}, though as noted above most hubs are not ICRs.  Similar results have been seen in other contexts, with those individuals in superordinate roles or rank in conversational contexts tending  to talk more \citep{fisek.et.al:ajs:1991} and higher ranking individuals being more likely to send and receive e-mail communications \citep{gibson.et.al:po:2019}.  It is thus highly plausible that ICRs are more attractive targets for interaction, and more active in initiating contact with others, thereby contributing to hub formation.  This may be particularly important in specialist networks, where centralizing coordination activity on a small number of pre-assigned hub roles is often a deliberate strategy (albeit one that may fail in the context of a real disaster).

\section*{Data and Methods}

\subsection*{Data}

\begin{table}
\centering
\setlength\tabcolsep{0.15cm}
\begin{tabular}{lrrrr}
 \hline
 & Actors & Events & \% ICR & Specialization \\ 
  \hline
Newark Maintenance &  27 &   77 & 3.70 & Non Spec. \\ 
  PATH Radio Comm &  32 &   70 & 6.25 & Non Spec. \\ 
  WTC Operations & 130 &  562 & 1.54 & Non Spec. \\ 
  Newark Operations Terminals & 138 & 1012 & 4.35 & Non Spec. \\ 
  PATH Control Desk & 229 & 1066 & 6.99 & Non Spec. \\ 
  Newark Facility Management & 237 & 1100 & 2.95 & Non Spec. \\ 
  WTC Vertical Trans & 246 &  780 & 1.22 & Non Spec. \\ 
  WTC Maintenance Electric & 256 &  864 & 6.25 & Non Spec. \\ 
  Newark Police &  24 &   83 & 8.33 & Specialist \\ 
  NJSPEN 2 &  32 &  149 & 15.62 & Specialist \\ 
  WTC Police &  37 &  481 & 8.11 & Specialist \\ 
  Newark CPD &  50 &  271 & 16.00 & Specialist \\ 
  PATH Police &  93 &  689 & 3.23 & Specialist \\ 
  Newark Command & 111 &  320 & 2.70 & Specialist \\ 
  WTC Security & 118 &  582 & 10.17 & Specialist \\ 
  NJSPEN 1 & 166 &  575 & 9.04 & Specialist \\ 
  Lincoln Tunnel Police & 229 & 1145 & 4.37 & Specialist \\ \hline
  Mean & 127 &  578 & 6.52 &  \\
   \hline
\end{tabular}
\vspace{10pt}
\caption{Summary statistics for the WTC radio networks. \label{tab_summary}}
\end{table}

The data we employ here are derived from transcripts of radio communications among responders to the WTC disaster on the morning of 9/11/2001, released by the Port Authority of New York and New Jersey and coded by \citet{butts.et.al:jms:2007}.  We analyze the radio communication transcripts from seventeen groups of responders to the WTC disaster, each of whom was using one radio channel exclusively. All transcripts begin immediately after the first airplane crashed into the North Tower at 8:46 am, and extend for three hours and 33 minutes or until communication was terminated by structural collapse (as occurred for some units located within the WTC complex).

\begin{figure}
\centering
\includegraphics[width=\textwidth]{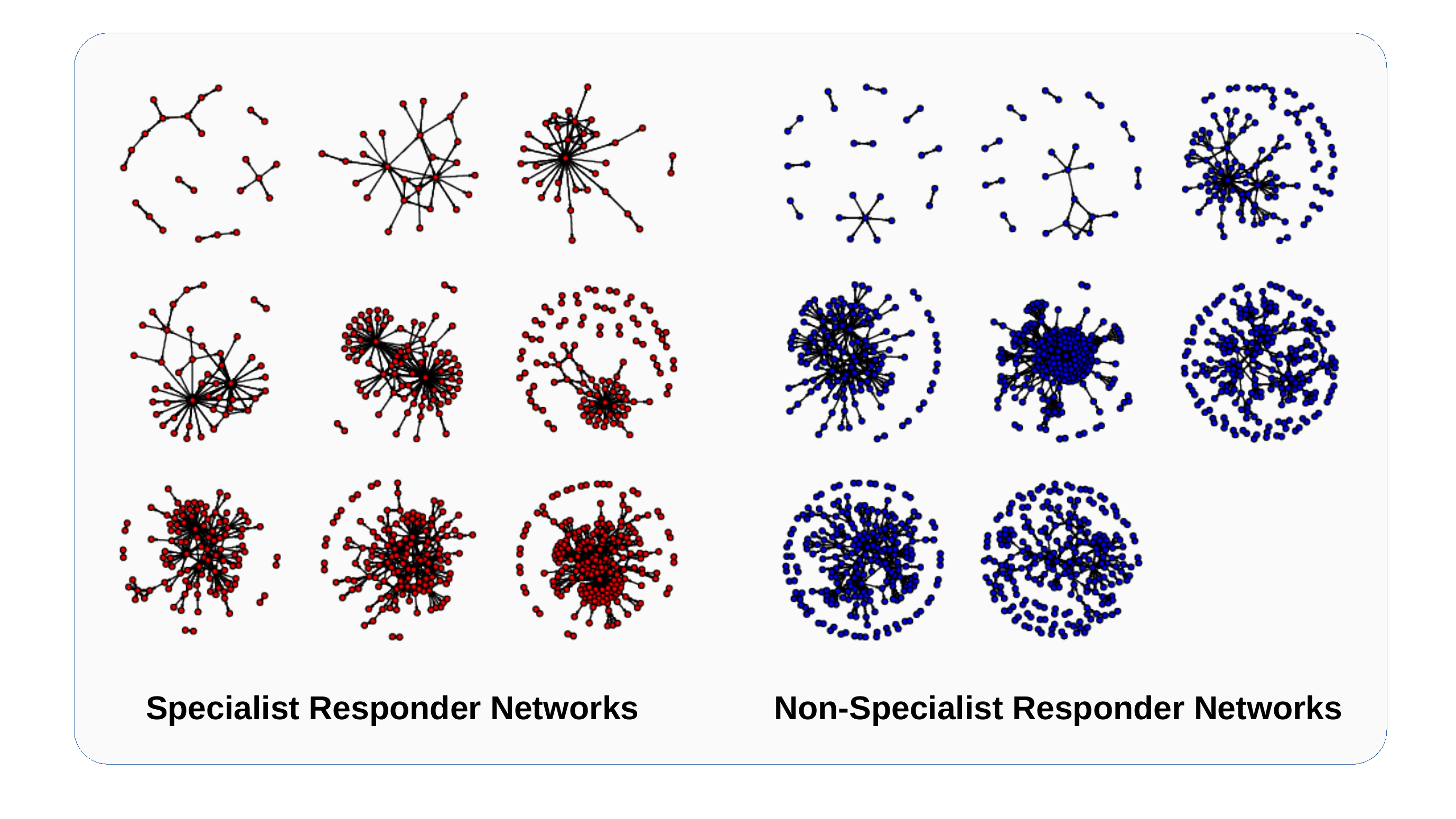}
\caption{Visualization of the time-marginalized WTC networks, sorted by specialization.  All networks show high levels of centralization, with a relatively small number of coordinators providing much of the connectivity in each case. \label{fig_radionets}}
\end{figure}

Based on information provided by the original transcriber and/or other transcript content, a unique identifier was assigned to the sender and named target(s) of each transmission \citep{butts.et.al:jms:2007}. Where one-to-many communications were encountered, each was coded as a series of dyadic transmissions from the sender to each of the named recipients (in the order named). Transmissions with no clear target(s), and/or targets that were identified only as a group (e.g., ``anyone,'' ``all units'') were not included. The resulting lists of ordered transmissions (one per transcript) comprised the relational event sets employed in subsequent analyses. Lengths range from 70 to 1,145 eligible transmissions, with the number of named communicants ranging from 24 to 256. Figure~\ref{fig_radionets} illustrates the networks in question, while showcasing the coordinative hub structure we aim to investigate. In conjunction with the relational event data itself, we consider the formal coordinative status of individual responders as an illustrative covariate, and distinguish between specialist and non-specialist responder networks, using the classification criteria provided by \citep{butts.et.al:jms:2007}.  A package containing the full data set is included as a supplement to this paper.

Summary statistics for the 17 networks are shown in Table~\ref{tab_summary}, including the number of individuals in the system (network size), the number of events, the fraction of individuals who are ICRs, and the specialist/non-specialist encoding. About 6.5\% of responders occupy ICRs, with this fraction varying from 1.2-16\%. \citet{butts.et.al:jms:2007} previously found that these networks are highly centralized, as is evident in Figure~\ref{fig_radionets}.

\subsection*{Methods}

\subsection*{Relational Event Models}

As in the original work by \citet{butts:sm:2008}, we use the relational event modeling framework to understand the effects of various conversational drivers involved in the formation of emergency communication networks. Because the 17 networks studied here vary greatly in size, specialization, location, and types of task performance, it is entirely plausible that different mechanisms are active in different groups; rather than attempting to fit a pooled model (or a hierarchical model with common effects, as in \citet{dubois.et.al:jmp:2013}), we instead proceed by (1) identifying a space of substantively plausible models, and then (2) performing complexity-penalized model selection (AICc) to find the effects active in each network.  We verify adequacy of the selected models using local prediction and recall.  (All models were fit and simulations performed using the \texttt{relevent} package \citep{butts:sw:2013} v1.1 for the R statistical computing system \citep{rteam:sw:2021}.)

The space of plausible mechanisms includes those discussed by \cite{butts:sm:2008}, as well as effects motivated by the background discussion above.  The effects considered include the following.  \emph{Preferential attachment}, where ego tends to call those with more airtime, is parameterized as the effect of normalized total communication volume on call receipt (\emph{NTDegRec}). This results in a single statistic whose parameter will be positive if preferential attachment is present and negative if a rotating turn effect is present.  Next, we consider cognitive effects: \emph{persistence} (previous out-alters are salient for ego's out-calls) is parameterized as the tendency for responders to direct calls to those that they have previously directed calls to, and \emph{recency} (more recent in-alters are salient for ego's out-calls) is parameterized as the tendency for responders to direct calls to those that they have most recently received calls from. A positive value for these parameters would indicate the presence of these cognitive mechanisms, while negative values might indicate a search pattern for individuals or perhaps an attempt to disseminate information across a wide swath of the network.

Similar to triadic effects in the static networks, there are potential patterns of triadic closure in the REM framework. For the first two of these statistics, one is parameterized as the number of outbound two paths from ego to some alter and the other as the number of inbound two paths from some alter to ego (\emph{OTPSnd} and \emph{ITPSnd}). For the former, if the parameter is positive, this indicates a pattern of transitive closure. Substantively, this may come from responders choosing to share information with individuals directly, rather than through third parties. If this parameter is negative, it may indicate that responders are choosing to rely on intermediaries to spread information throughout the network. For the latter, a positive coefficient for this statistic would indicate a tendency toward cyclic closure. This would indicate that if individuals are relaying information through intermediaries, those on the receiving end have a tendency to reply directly. A negative coefficient would indicate a tendency away from cycles and imply that individuals more often respond to the intermediaries, rather than those two steps away. We also include out-bound and incoming shared partner effects \emph{OSPSnd} and \emph{ISPSnd} - analogous to the \emph{sibling effects} of \citet{fararo.sunshine:bk:1964} - to help understand other triadic communication patterns between communicants.

The next set of parameters are \emph{conversational inertia terms} or \emph{participation shifts} (tendencies reflecting ``local'' conversational norms \citep[from ][]{gibson:sf:2003}).  The ones included in the following analysis are \emph{PSAB-BA} or call and response, \emph{PSAB-AY} and \emph{PSAB-XB} or persistence of source and target, respectively, \emph{PSAB-XA} or source attraction, and finally \emph{PSAB-BY} or ``handing off'' of communication. Lastly, the pre-disaster organizational role is parameterized as a covariate effect for ego occupying and institutional coordinative role (\emph{ICR}) on in and out calls.  

To estimate model parameters, we perform Bayesian inference using the Laplace approximation on all 17 WTC networks, using diffuse $t$-priors with prior location 0, scale 10, and 4 degrees of freedom; model selection is performed by AICc minimization, as described below. Comparison of parameters from fitted models is used to assess the differences between specialist and non-specialist responders, similar versus dissimilar mechanisms driving behavior across networks, and to evaluate the respective impact of ICR membership and endogenous factors in shaping the emergence of coordinator roles.

\subsection*{Model Selection}

We begin by noting that it is plausible that different mechanisms may be active in different networks, and that - recognizing that any model for a complex system is at best an approximation, rather than a ``true'' model - we focus on identifying collections of mechanisms likely to have good generalization performance (in terms of deviance under hypothetical replication).  This motivates the use of model selection via the (sample-size corrected) Akaike's Information Criterion (AICc), which is a complexity-adjusted estimator of replicate deviance \citep{akaike:itac:1974,bozdogan:jmp:2000}.  Given the set of candidate terms, we attempt to select the model that minimizes the AICc (a procedure analogous to so-called ``$L_0$'' regularized model selection).  This is a non-convex optimization problem, requiring exhaustive enumeration for an exact solution.  We thus approximate the exact solution by a local optimization procedure (aka ``hill-climbing'') that seeks the AICc-optimal model by steepest descent.  At each iteration, the procedure considers all single-term changes to the current model (addition or deletion of terms), taking the change that results in the greatest AICc reduction and terminating when no improvement can be made.  We initialize the search procedure with the empty (null) model.  As a check on the quality of the hill climbing optimization process, we generated a full factorial design using the smallest network, PATH Radio Communications, which was small enough for enumerative search; the hill-climbing procedure was indeed able to find the optimal model in this case.  Manual inspection of paths followed by the search procedure likewise showed no evidence of pathological behavior.  While - as with any heuristic optimization procedure - it is not possible to guarantee that the optimal model is identified in every case, this approach does guarantee that (1) the selected model is locally optimal (i.e., it cannot be improved by adding or removing effects), and (2) if not a null or one-term model, the selected model is better than the null or any one-term model.  

\subsection*{Knock-Out Simulation Experiment}

To better understand how preferential attachment, ICR, and conversational inertia terms influence the hub-generation process in these radio communication networks, we perform a series of computational ``knock-out'' experiments for these effects.  Specifically, we use the \texttt{simulate} function from the \texttt{relevent} package to take draws from the posterior predictive distribution of relational event trajectories from each network, drawing 50 trajectories of equal length to the observed data (and with identical covariates) with parameters simulated using the fitted posterior mean and variance-covariance matrix.  (We employ the Laplace approximation, treating the posterior as multivariate Gaussian.)  We then conduct 50 replicate simulations for each of four knock-out conditions: in the first we remove (set to 0) the preferential attachment term (NTDegRec); in the second we remove conversational inertia terms (e.g., PSAB-BA, PSAB-BY etc.); in the third we remove the ICR term; and the final condition all of the three hub-forming term sets are removed. All non-removed parameters in each condition take the same values as were used for the full model.  This process results in a total of 5100 simulated trajectories across the 17 networks.

In order to examine the impact of mechanisms on the generalized tendency to form hubs, we need a way to characterize how concentrated the distribution of the communication volume is. We choose a widely used concentration index, the Theil index \citep{theil:bk:1967} for this purpose, which is has been used to study various phenomena: from income inequality \citep{silva.leichenko:eg:2009}, crime \citep{kang:jpe:2016}, to health disparities \citep{borrell.talih:sim:2011,manz.mansmann:plos:2021}, just to name a few. It has a natural interpretation in terms of the entropic “cost” of going from an equal distribution to that of the observed system -- we can think of the Theil index as expressing  the extent to which social mechanisms are systematically organizing/biasing activity, versus letting it happen at random, making it a sensible measure of communication volume concentration in our networks. 

Using the simulations in each condition, we then calculate the Theil index for each network using total communication volume per actor (taking the mean index value over all simulated replicates). We assess the contribution of each mechanism to the extent of hub formation by examining the reduction in the mean Theil index when the respective set of terms is removed (versus the full model).  Larger reductions imply a greater role for the associated mechanism in hub formation in the respective network.  (For cases in which a given mechanism was not in the best fitting model, its contribution is trivially 0.)

\section*{Results}

\subsection*{Model Adequacy}

\begin{table}
\centering
\setlength\tabcolsep{0.15cm}
\begin{tabular}{lrrrrrrrr}
  \hline
          & \multicolumn{4}{c}{Match to Next Event} & & \multicolumn{3}{c}{Recall (Coverage)} \\
          & \multicolumn{2}{c}{Either Match} & \multicolumn{2}{c}{Both Match} && Top & Top & Top \\
  Network & Fitted & Null & Fitted & Null && 1\% & 5\% & 10\% \\  \hline
PATH Radio Communications & 0.67 & 0.06 & 0.56 & 0.001 && 0.67 & 0.73 & 0.86 \\
  Lincoln Tunnel Police & 0.23 & 0.01 & 0.08 & $<$0.001 && 0.70 & 0.81 & 0.86 \\
  Newark Command & 0.72 & 0.02 & 0.64 & $<$0.001 && 0.76 & 0.83 & 0.90 \\
  Newark Police & 0.81 & 0.08 & 0.75 & 0.002 && 0.82 & 0.83 & 0.88 \\
  Newark CPD & 0.72 & 0.04 & 0.56 & $<$0.001 && 0.74 & 0.86 & 0.91 \\
  Newark Operations Terminals & 0.75 & 0.01 & 0.65 & $<$0.001 && 0.80 & 0.90 & 0.93 \\
  Newark Maintenance & 0.79 & 0.07 & 0.78 & 0.001 && 0.87 & 0.88 & 0.90 \\
  PATH Control Desk & 0.75 & 0.01 & 0.62 & $<$0.001 && 0.83 & 0.91 & 0.94 \\
  NJSPEN 1 & 0.61 & 0.01 & 0.49 & $<$0.001 && 0.67 & 0.80 & 0.83 \\
  NJSPEN 2 & 0.65 & 0.06 & 0.50 & 0.001 && 0.65 & 0.79 & 0.89 \\
  WTC Operations & 0.68 & 0.02 & 0.56 & $<$0.001 && 0.77 & 0.90 & 0.92 \\
  WTC Police & 0.79 & 0.05 & 0.68 & $<$0.001 && 0.82 & 0.93 & 0.96 \\
  WTC Vertical Transportation & 0.64 & 0.01 & 0.56 & $<$0.001 && 0.73 & 1.00 & 1.00 \\
  Newark Facility Management & 0.72 & 0.01 & 0.67 & $<$0.001 && 0.78 & 0.87 & 0.89 \\
  PATH Police & 0.78 & 0.02 & 0.62 & $<$0.001 && 0.83 & 0.96 & 0.97 \\
  WTC Security & 0.71 & 0.02 & 0.59 & $<$0.001 && 0.74 & 0.87 & 0.92 \\
  WTC Maintenance Electric & 0.57 & 0.01 & 0.53 & $<$0.001 && 0.76 & 0.82 & 0.85 \\
  \hline
  Mean & 0.68 & 0.03 & 0.58 & $<$0.001 && 0.76 & 0.86 & 0.91 \\
   \hline
\end{tabular}
\vspace{10pt}
\caption{Model adequacy checks.  Observed and null probabilities of matching features of the next event in each sequence (``Eiher'' implies that sender or receiver match, while ``All'' implies that both sender and receiver match). All models correctly identify the next event with probability greatly exceeding the null model, on average doing so the majority of the time. Recall columns show the fraction of observed events covered by the respective fraction of probability-ordered predictions (higher is better). \label{tab_gof}}

\end{table}

In order to check the adequacy of our models, we  assess the ability of each model to accurately identify the next event in its respective event history. Table~\ref{tab_gof} shows the observed probability that the model was able to accurately predict either sender or receiver for a given event, followed by the null random probability of predicting either sender or receiver, then the observed probability of the model correctly guessing both the sender \emph{and} receiver of a given event, and finally, its corresponding null random choice probability. Despite the seemingly chaotic communication environment of the WTC, the models perform extremely well.  On average across all models, our ability to predict either of the two individuals in a given communication event correctly is 68\%, meaning that in approximately 68\% of cases the model's top choice for the next event in the sequence specifies the sender or receiver correctly.  The lowest prediction probability for a specific model is for the largest network of the set, Lincoln Tunnel, and even in this case the model is able to predict either the sender or receiver from the set of 229 individuals 23\% of the time. This is substantially larger than the probability of correct prediction under the null model.  We also examine the probability of correctly predicting both sender \emph{and} receiver, an extremely difficult task.  On average, we find that our models can accurately predict both sender and receiver for the next event 58\% of the time, with for some models as high as 78\% and for the Lincoln Tunnel model, 8\% of the time.  While the latter seems low in absolute terms, we observe that the likelihood of randomly guessing the sender and receiver combination for Lincoln Tunnel is 0.000002\%, with 10 other networks also having random predictions correct at a similarly low probability.  While getting the next event right is a very strong test of adequacy, we also consider more general recall rates, i.e., the fraction of observed events that are ``covered'' by the top $k$\% of predictions (the events judged most likely to occur).  We examine recall rates for the top 1\%, 5\%, and 10\% of predictions, providing a sense of the ability of the model to focus attention on events that are relatively likely to occur.  As Table~\ref{tab_gof} shows, the vast majority of events are covered by the top few percent of predictions, with 76\% on average contained within the top 1\%, and over 90\% within the top 10\%.  These results suggest that the selected models are able to capture the dynamics of the WTC system with sufficient fidelity to advance to the next stage of analysis.

\subsection*{Relational Event Model Analysis}

\begin{figure}
\centering
\includegraphics[width=\textwidth]{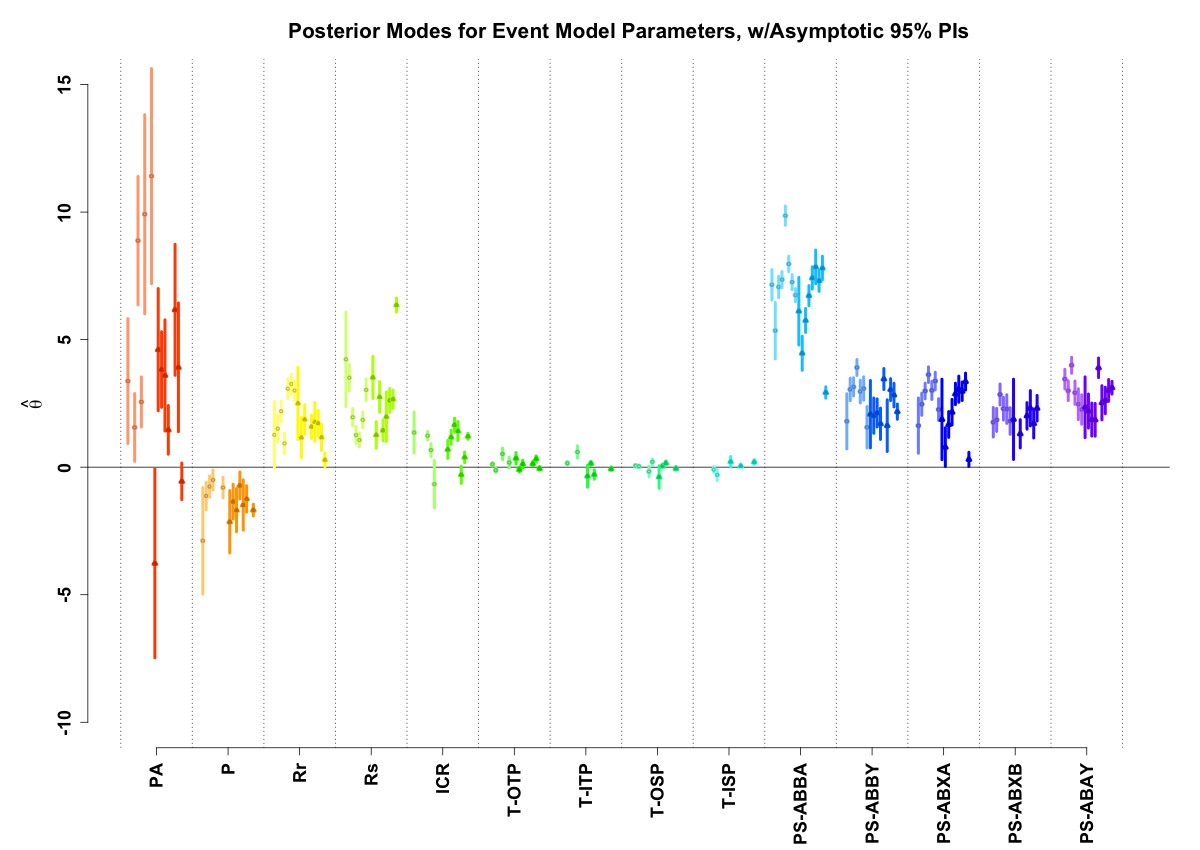}
\caption{Posterior modes and their asymptotic 95\% posterior intervals for the event model parameters for all 17 WTC radio communication networks. Darker colored line segments represent specialist networks, while lightly colored line segments represent non-specialist networks.  Term codes are as follows: PA, preferential attachment; P, persistence; Rr, recency of receipt; Rs, recency of sending; ICR, role effect; T-OTP, outgoing two-path; I-ITP, incoming two-path; T-ISP, incoming shared partners; T-OSP, outgoing shared partners; PS-$*$, P-shift effects. \label{fig_coef}}
\end{figure}

The coefficients for the selected models (posterior mean estimates) are shown in Tables~\ref{tab_models_1}--\ref{tab_models_2}; 95\% posterior intervals can also be seen in Figure~\ref{fig_coef}. For convenience in identifying effects whose signs are well-determined by the data, Tables~\ref{tab_models_1} and \ref{tab_models_2} identify posterior mean estimates for which the central 95\%, 99\%, and 99.9\% posterior intervals (sometimes called ``credible intervals'') exclude 0.  %

\begin{table}
\centering
\resizebox{\columnwidth}{!}{%
\begin{tabular}{@{\extracolsep{7pt}}lp{1.6cm}p{1.6cm}p{1.7cm}p{1.4cm}p{1.6cm}p{1.5cm}p{1.7cm}p{1.6cm}p{1.1cm}}
\\
  \cline{1-10}
 & Lincoln & Newark & Newark & Newark & NJSPEN & NJSPEN & WTC & PATH & WTC \\ 
& Tunnel & Command & Police & CPD & 1 & 2 & Police & Police & Security\\
  \cline{1-10}
PA & $-$0.56 &  & $-$3.77$^{*}$ & 3.60$^{**}$ & 3.91$^{**}$ & 4.61$^{***}$ & 3.83$^{***}$ & 1.46$^{**}$ & 6.17$^{***}$ \\ 
  & (0.37) &  & (1.89) & (1.11) & (1.29) & (1.22) & (0.76) & (0.49) & (1.31) \\ 
  & & & & & & & & & \\ 
 P & $-$1.68$^{***}$ & $-$1.48$^{**}$ &  & $-$1.69$^{***}$ &  & $-$2.14$^{***}$ & $-$1.35$^{***}$ & $-$0.72$^{**}$ & $-$1.24$^{***}$ \\ 
  & (0.12) & (0.50) &  & (0.43) &  & (0.62) & (0.35) & (0.27) & (0.27) \\ 
  & & & & & & & & & \\ 
 Rr & 0.29$^{*}$ & 1.77$^{***}$ & 2.50$^{***}$ &  & 1.17$^{***}$ & 1.16$^{**}$ & 1.87$^{***}$ & 1.59$^{***}$ & 1.72$^{***}$ \\ 
  & (0.14) & (0.39) & (0.72) &  & (0.27) & (0.40) & (0.29) & (0.23) & (0.26) \\ 
  & & & & & & & & & \\ 
 Rs & 6.36$^{***}$ & 1.98$^{***}$ &  & 2.75$^{***}$ & 2.66$^{***}$ & 3.52$^{***}$ & 1.26$^{***}$ & 1.44$^{***}$ & 2.62$^{***}$ \\ 
  & (0.14) & (0.50) &  & (0.31) & (0.18) & (0.42) & (0.27) & (0.21) & (0.24) \\ 
  & & & & & & & & & \\ 
 ICR & 1.21$^{***}$ & 1.42$^{***}$ &  & 1.17$^{***}$ & 0.39$^{***}$ &  & 0.69$^{***}$ & 1.66$^{***}$ & $-$0.30 \\ 
  & (0.06) & (0.19) &  & (0.14) & (0.11) &  & (0.18) & (0.13) & (0.17) \\ 
  & & & & & & & & & \\ 
 T-OTP & $-$0.05 &  &  & 0.12 & 0.33$^{***}$ & 0.35$^{**}$ & $-$0.09 &  & 0.14$^{**}$ \\ 
  & (0.03) &  &  & (0.08) & (0.06) & (0.11) & (0.05) &  & (0.04) \\ 
  & & & & & & & & & \\ 
 T-ITP & $-$0.07$^{**}$ &  &  & $-$0.29$^{**}$ &  & $-$0.35 & 0.15$^{***}$ &  &  \\ 
  & (0.03) &  &  & (0.09) &  & (0.22) & (0.04) &  &  \\ 
  & & & & & & & & & \\ 
 T-OSP &  &  &  & 0.16$^{***}$ &  & $-$0.38 & 0.04$^{*}$ &  & $-$0.05 \\ 
  &  &  &  & (0.04) &  & (0.23) & (0.02) &  & (0.03) \\ 
  & & & & & & & & & \\
 T-ISP & 0.20$^{***}$ &  &  &  &  & 0.22$^{*}$ &  & 0.05$^{***}$ &  \\ 
  & (0.05) &  &  &  &  & (0.10) &  & (0.01) &  \\ 
  & & & & & & & & & \\  
 PS-ABBA & 2.93$^{***}$ & 7.85$^{***}$ & 6.11$^{***}$ & 6.72$^{***}$ & 7.81$^{***}$ & 4.47$^{***}$ & 5.75$^{***}$ & 7.42$^{***}$ & 7.30$^{***}$ \\ 
  & (0.11) & (0.34) & (0.68) & (0.20) & (0.24) & (0.34) & (0.24) & (0.22) & (0.21) \\ 
  & & & & & & & & & \\ 
 PS-ABBY & 2.18$^{***}$ & 1.63$^{**}$ & 2.08$^{**}$ & 1.70$^{***}$ & 2.83$^{***}$ & 2.02$^{***}$ & 2.13$^{***}$ & 3.46$^{***}$ & 3.04$^{***}$ \\ 
  & (0.15) & (0.52) & (0.67) & (0.31) & (0.24) & (0.35) & (0.28) & (0.21) & (0.22) \\ 
  & & & & & & & & & \\ 
 PS-ABXA & 0.31$^{*}$ & 3.06$^{***}$ & 1.87$^{*}$ & 2.16$^{***}$ & 3.35$^{***}$ & 0.79$^{*}$ & 1.67$^{***}$ & 2.87$^{***}$ & 2.98$^{***}$ \\ 
  & (0.14) & (0.26) & (0.81) & (0.26) & (0.17) & (0.39) & (0.26) & (0.21) & (0.19) \\ 
  & & & & & & & & & \\ 
 PS-ABXB &  & 2.31$^{***}$ & 1.88$^{*}$ &  & 2.31$^{***}$ &  & 1.31$^{***}$ & 2.01$^{***}$ & 1.71$^{***}$ \\ 
  &  & (0.35) & (0.80) &  & (0.26) &  & (0.28) & (0.27) & (0.29) \\ 
  & & & & & & & & & \\ 
 PS-ABAY & 3.11$^{***}$ & 2.52$^{***}$ & 2.35$^{***}$ & 1.86$^{***}$ & 3.01$^{***}$ & 2.20$^{***}$ & 1.87$^{***}$ & 3.90$^{***}$ & 2.62$^{***}$ \\ 
  & (0.13) & (0.34) & (0.61) & (0.33) & (0.22) & (0.34) & (0.33) & (0.20) & (0.27) \\ 
  & & & & & & & & & \\ 
\cline{1-10} \\ 
AICc & 16,368.72 & 2,295.98 & 344.67 & 1,791.26 & 5,943.17 & 979.30 & 2,176.21 & 4,137.43 & 4,562.27 \\ 
 
\cline{1-10}
\multicolumn{10}{c}{Stars indicate 0 excluded from posterior intervals: 95\%: $^*$, 99\%: $^{**}$, 99.9\%: $^{***}$} \\ \hline
\end{tabular}
}
\vspace{10pt}
\caption{Posterior means and standard deviations for AICc-selected models for the specialist networks. \label{tab_models_1}}
\end{table}

\begin{table}
\centering
\resizebox{\columnwidth}{!}{%
\begin{tabular}{@{\extracolsep{8pt}}lp{1.1cm}p{1.5cm}p{1.7cm}p{1.2cm}p{1.5cm}p{1.4cm}p{1.7cm}p{1.5cm}}
  \\
  \cline{1-9}
 & PATH & Newark & Newark & PATH & WTC  & WTC & Newark & WTC\\
 & Radio & Operations& Maintenance & Control & Operations & Vertical & Facility & Maintenance\\
 & Comm & Terminals &  & Desk & & Transport & Management & Electric\\
  \cline{1-9}
PA &  & 8.88$^{***}$ & 3.37$^{**}$ & 2.56$^{***}$ & 1.56$^{*}$ &  & 9.92$^{***}$ & 11.41$^{***}$ \\ 
  &  & (1.28) & (1.25) & (0.50) & (0.68) &  & (1.99) & (2.15) \\ 
  & & & & & & & & \\ 
 P & $-$2.88$^{**}$ & $-$0.76$^{***}$ &  & $-$0.50$^{*}$ & $-$1.13$^{***}$ &  &  & $-$0.80$^{***}$ \\ 
  & (1.06) & (0.21) &  & (0.20) & (0.28) &  &  & (0.21) \\ 
  & & & & & & & & \\ 
 Rr & 1.27 & 2.19$^{***}$ &  & 0.93$^{***}$ & 1.51$^{***}$ & 3.26$^{***}$ & 3.07$^{***}$ & 3.00$^{***}$ \\ 
  & (0.67) & (0.20) &  & (0.21) & (0.27) & (0.20) & (0.20) & (0.19) \\ 
  & & & & & & & & \\ 
 Rs & 4.23$^{***}$ & 1.96$^{***}$ &  & 1.26$^{***}$ & 3.51$^{***}$ & 1.86$^{***}$ & 1.06$^{***}$ & 3.03$^{***}$ \\ 
  & (0.95) & (0.17) &  & (0.17) & (0.27) & (0.16) & (0.14) & (0.22) \\ 
  & & & & & & & & \\ 
ICR &  &  & 1.36$^{***}$ & 1.23$^{***}$ &  & $-$0.66 & 0.67$^{***}$ &  \\ 
  &  &  & (0.41) & (0.09) &  & (0.48) & (0.13) &  \\ 
  & & & & & & & & \\ 
 T-OTP &  & $-$0.12$^{**}$ &  &  & 0.11$^{***}$ &  & 0.52$^{***}$ & 0.18 \\ 
  &  & (0.04) &  &  & (0.03) &  & (0.12) & (0.11) \\ 
  & & & & & & & & \\ 
 T-ITP &  & 0.16$^{***}$ &  &  &  & 0.60$^{***}$ &  &  \\ 
  &  & (0.03) &  &  &  & (0.13) &  &  \\ 
  & & & & & & & & \\ 
 T-OSP &  & 0.04$^{***}$ &  &  & 0.06$^{*}$ & $-$0.17 &  & 0.21$^{***}$ \\ 
  &  & (0.01) &  &  & (0.03) & (0.11) &  & (0.06) \\ 
  & & & & & & & & \\ 
 T-ISP &  &  &  & $-$0.09 &  &  & $-$0.30$^{**}$ &  \\ 
  &  &  &  & (0.06) &  &  & (0.12) &  \\ 
  & & & & & & & & \\ 
 PS-ABBA & 5.35$^{***}$ & 7.35$^{***}$ & 7.15$^{***}$ & 9.86$^{***}$ & 7.07$^{***}$ & 7.25$^{***}$ & 7.97$^{***}$ & 6.74$^{***}$ \\ 
  & (0.57) & (0.16) & (0.30) & (0.19) & (0.22) & (0.15) & (0.16) & (0.13) \\ 
  & & & & & & & & \\ 
 PS-ABBY & 1.80$^{**}$ & 3.15$^{***}$ &  & 3.91$^{***}$ & 3.05$^{***}$ & 3.08$^{***}$ & 2.96$^{***}$ & 1.57$^{***}$ \\ 
  & (0.55) & (0.18) &  & (0.16) & (0.23) & (0.23) & (0.23) & (0.41) \\ 
  & & & & & & & & \\ 
 PS-ABXA & 1.63$^{**}$ & 2.99$^{***}$ &  & 3.63$^{***}$ & 2.47$^{***}$ & 3.38$^{***}$ & 3.01$^{***}$ & 2.26$^{***}$ \\ 
  & (0.55) & (0.16) &  & (0.16) & (0.22) & (0.18) & (0.18) & (0.22) \\ 
  & & & & & & & & \\ 
 PS-ABXB &  & 1.85$^{***}$ &  & 2.85$^{***}$ & 1.76$^{***}$ & 2.28$^{***}$ & 2.29$^{***}$ & 1.81$^{***}$ \\ 
  &  & (0.23) &  & (0.21) & (0.30) & (0.29) & (0.23) & (0.27) \\ 
  & & & & & & & & \\ 
 PS-ABAY &  & 3.00$^{***}$ &  & 3.99$^{***}$ & 3.45$^{***}$ & 2.47$^{***}$ & 2.92$^{***}$ & 2.26$^{***}$ \\ 
  &  & (0.20) &  & (0.16) & (0.20) & (0.31) & (0.23) & (0.30) \\ 
  & & & & & & & & \\  
\cline{1-9} \\ 
AICc & 483.66 & 6,862.60 & 289.70 & 8,336.29 & 4,545.03 & 7,584.97 & 8,382.05 & 8,807.99 \\ 
\cline{1-9}
\multicolumn{9}{c}{Stars indicate 0 Excluded from posterior intervals: 95\%: $^*$, 99\%: $^{**}$, 99.9\%: $^{***}$} \\ \hline
\end{tabular}
}
\vspace{10pt}
\caption{Posterior means and standard deviations for AICc-selected models for the non-specialist networks. \label{tab_models_2}}
\end{table}

To summarize the mechanisms active in each network, Table~\ref{tab_effectincl} provides a schematic view of effects across networks, with their estimated signs.  One of the more striking results is that all network final models contain the turn taking P-shift term, PSAB-BA, meaning that in all sizes of networks and regardless of specialization, turn taking is an important feature in driving the local patterns interaction. Additionally, the PSAB-BA term is the only one present in all final models. Some other commonly occurring terms that do not relate to the hub formation process, but that modulate the overall communication process and control for communicational differences between networks include: recency in receiving (RRecSnd), present in 88.2\% of all networks, recency in sending (RSndSnd) occurring in 88.2\% of all networks, persistence (FrPSndSnd), which is in 70.6\% of all networks, and the outbound two path term (OTPSnd), present in 58.8\% of all final networks. Other effects, like OSPSnd (present in 47\% of all networks), ITPnd (present in 35.3\% of all networks), and ISPnd (present in 29.4\% of all networks), are less common and have little effect.

\definecolor{lred}{cmyk}{0,.47,.51,.03}
\definecolor{lgreen }{cmyk}{.51,0,.47,.03}
\begin{table}
\centering
\resizebox{\columnwidth}{!}{%
\setlength\tabcolsep{0.06cm}
\begin{tabular}{lcccccccccccccc}
  \hline
 & P & Rr & Rs & T-OTP & T-ITP & T-OSP & T-ISP & PS-ABBA & PS-ABAY & PS-ABXB & PS-ABXA & PS-ABBY & PA & ICR \\
  \hline
PATH Radio Comm & \cellcolor{lred} \textbf{-}  & \cellcolor{lgreen } \textbf{+}  & \cellcolor{lgreen } \textbf{+}  &  &  &  &  & \cellcolor{lgreen } \textbf{+}  &  &  & \cellcolor{lgreen } \textbf{+}  & \cellcolor{lgreen } \textbf{+}  &  &  \\
  Lincoln Tunnel Police & \cellcolor{lred} \textbf{-}  & \cellcolor{lgreen } \textbf{+}  & \cellcolor{lgreen } \textbf{+}  & \cellcolor{lred} \textbf{-}  & \cellcolor{lred} \textbf{-}  &  & \cellcolor{lgreen } \textbf{+}  & \cellcolor{lgreen } \textbf{+}  & \cellcolor{lgreen } \textbf{+}  &  & \cellcolor{lgreen } \textbf{+}  & \cellcolor{lgreen } \textbf{+}  & \cellcolor{lred} \textbf{-}  & \cellcolor{lgreen } \textbf{+}  \\
  Newark Command & \cellcolor{lred} \textbf{-}  & \cellcolor{lgreen } \textbf{+}  & \cellcolor{lgreen } \textbf{+}  &  &  &  &  & \cellcolor{lgreen } \textbf{+}  & \cellcolor{lgreen } \textbf{+}  & \cellcolor{lgreen } \textbf{+}  & \cellcolor{lgreen } \textbf{+}  & \cellcolor{lgreen } \textbf{+}  &  & \cellcolor{lgreen } \textbf{+}  \\
  Newark Police &  & \cellcolor{lgreen } \textbf{+}  &  &  &  &  &  & \cellcolor{lgreen } \textbf{+}  & \cellcolor{lgreen } \textbf{+}  & \cellcolor{lgreen } \textbf{+}  & \cellcolor{lgreen } \textbf{+}  & \cellcolor{lgreen } \textbf{+}  & \cellcolor{lred} \textbf{-}  &  \\
  Newark CPD & \cellcolor{lred} \textbf{-}  &  & \cellcolor{lgreen } \textbf{+}  & \cellcolor{lgreen } \textbf{+}  & \cellcolor{lred} \textbf{-}  & \cellcolor{lgreen } \textbf{+}  &  & \cellcolor{lgreen } \textbf{+}  & \cellcolor{lgreen } \textbf{+}  &  & \cellcolor{lgreen } \textbf{+}  & \cellcolor{lgreen } \textbf{+}  & \cellcolor{lgreen } \textbf{+}  & \cellcolor{lgreen } \textbf{+}  \\
  Newark Operations Terminals & \cellcolor{lred} \textbf{-}  & \cellcolor{lgreen } \textbf{+}  & \cellcolor{lgreen } \textbf{+}  & \cellcolor{lred} \textbf{-}  & \cellcolor{lgreen } \textbf{+}  & \cellcolor{lgreen } \textbf{+}  &  & \cellcolor{lgreen } \textbf{+}  & \cellcolor{lgreen } \textbf{+}  & \cellcolor{lgreen } \textbf{+}  & \cellcolor{lgreen } \textbf{+}  & \cellcolor{lgreen } \textbf{+}  & \cellcolor{lgreen } \textbf{+}  &  \\
  Newark Maintenance &  &  &  &  &  &  &  & \cellcolor{lgreen } \textbf{+}  &  &  &  &  & \cellcolor{lgreen } \textbf{+}  & \cellcolor{lgreen } \textbf{+}  \\
  PATH Control Desk & \cellcolor{lred} \textbf{-}  & \cellcolor{lgreen } \textbf{+}  & \cellcolor{lgreen } \textbf{+}  &  &  &  & \cellcolor{lred} \textbf{-}  & \cellcolor{lgreen } \textbf{+}  & \cellcolor{lgreen } \textbf{+}  & \cellcolor{lgreen } \textbf{+}  & \cellcolor{lgreen } \textbf{+}  & \cellcolor{lgreen } \textbf{+}  & \cellcolor{lgreen } \textbf{+}  & \cellcolor{lgreen } \textbf{+}  \\
  NJSPEN 1 &  & \cellcolor{lgreen } \textbf{+}  & \cellcolor{lgreen } \textbf{+}  & \cellcolor{lgreen } \textbf{+}  &  &  &  & \cellcolor{lgreen } \textbf{+}  & \cellcolor{lgreen } \textbf{+}  & \cellcolor{lgreen } \textbf{+}  & \cellcolor{lgreen } \textbf{+}  & \cellcolor{lgreen } \textbf{+}  & \cellcolor{lgreen } \textbf{+}  & \cellcolor{lgreen } \textbf{+}  \\
  NJSPEN 2 & \cellcolor{lred} \textbf{-}  & \cellcolor{lgreen } \textbf{+}  & \cellcolor{lgreen } \textbf{+}  & \cellcolor{lgreen } \textbf{+}  & \cellcolor{lred} \textbf{-}  & \cellcolor{lred} \textbf{-}  & \cellcolor{lgreen } \textbf{+}  & \cellcolor{lgreen } \textbf{+}  & \cellcolor{lgreen } \textbf{+}  &  & \cellcolor{lgreen } \textbf{+}  & \cellcolor{lgreen } \textbf{+}  & \cellcolor{lgreen } \textbf{+}  &  \\
  WTC Operations & \cellcolor{lred} \textbf{-}  & \cellcolor{lgreen } \textbf{+}  & \cellcolor{lgreen } \textbf{+}  & \cellcolor{lgreen } \textbf{+}  &  & \cellcolor{lgreen } \textbf{+}  &  & \cellcolor{lgreen } \textbf{+}  & \cellcolor{lgreen } \textbf{+}  & \cellcolor{lgreen } \textbf{+}  & \cellcolor{lgreen } \textbf{+}  & \cellcolor{lgreen } \textbf{+}  & \cellcolor{lgreen } \textbf{+}  &  \\
  WTC Police & \cellcolor{lred} \textbf{-}  & \cellcolor{lgreen } \textbf{+}  & \cellcolor{lgreen } \textbf{+}  & \cellcolor{lred} \textbf{-}  & \cellcolor{lgreen } \textbf{+}  & \cellcolor{lgreen } \textbf{+}  &  & \cellcolor{lgreen } \textbf{+}  & \cellcolor{lgreen } \textbf{+}  & \cellcolor{lgreen } \textbf{+}  & \cellcolor{lgreen } \textbf{+}  & \cellcolor{lgreen } \textbf{+}  & \cellcolor{lgreen } \textbf{+}  & \cellcolor{lgreen } \textbf{+}  \\
  WTC Vertical Trans &  & \cellcolor{lgreen } \textbf{+}  & \cellcolor{lgreen } \textbf{+}  &  & \cellcolor{lgreen } \textbf{+}  & \cellcolor{lred} \textbf{-}  &  & \cellcolor{lgreen } \textbf{+}  & \cellcolor{lgreen } \textbf{+}  & \cellcolor{lgreen } \textbf{+}  & \cellcolor{lgreen } \textbf{+}  & \cellcolor{lgreen } \textbf{+}  &  & \cellcolor{lred} \textbf{-}  \\
  Newark Facility Management &  & \cellcolor{lgreen } \textbf{+}  & \cellcolor{lgreen } \textbf{+}  & \cellcolor{lgreen } \textbf{+}  &  &  & \cellcolor{lred} \textbf{-}  & \cellcolor{lgreen } \textbf{+}  & \cellcolor{lgreen } \textbf{+}  & \cellcolor{lgreen } \textbf{+}  & \cellcolor{lgreen } \textbf{+}  & \cellcolor{lgreen } \textbf{+}  & \cellcolor{lgreen } \textbf{+}  & \cellcolor{lgreen } \textbf{+}  \\
  PATH Police & \cellcolor{lred} \textbf{-}  & \cellcolor{lgreen } \textbf{+}  & \cellcolor{lgreen } \textbf{+}  &  &  &  & \cellcolor{lgreen } \textbf{+}  & \cellcolor{lgreen } \textbf{+}  & \cellcolor{lgreen } \textbf{+}  & \cellcolor{lgreen } \textbf{+}  & \cellcolor{lgreen } \textbf{+}  & \cellcolor{lgreen } \textbf{+}  & \cellcolor{lgreen } \textbf{+}  & \cellcolor{lgreen } \textbf{+}  \\
  WTC Security & \cellcolor{lred} \textbf{-}  & \cellcolor{lgreen } \textbf{+}  & \cellcolor{lgreen } \textbf{+}  & \cellcolor{lgreen } \textbf{+}  &  & \cellcolor{lred} \textbf{-}  &  & \cellcolor{lgreen } \textbf{+}  & \cellcolor{lgreen } \textbf{+}  & \cellcolor{lgreen } \textbf{+}  & \cellcolor{lgreen } \textbf{+}  & \cellcolor{lgreen } \textbf{+}  & \cellcolor{lgreen } \textbf{+}  & \cellcolor{lred} \textbf{-}  \\
  WTC Maintenance Electric & \cellcolor{lred} \textbf{-}  & \cellcolor{lgreen } \textbf{+}  & \cellcolor{lgreen } \textbf{+}  & \cellcolor{lgreen } \textbf{+}  &  & \cellcolor{lgreen } \textbf{+}  &  & \cellcolor{lgreen } \textbf{+}  & \cellcolor{lgreen } \textbf{+}  & \cellcolor{lgreen } \textbf{+}  & \cellcolor{lgreen } \textbf{+}  & \cellcolor{lgreen } \textbf{+}  & \cellcolor{lgreen } \textbf{+}  &  \\
   \hline
\end{tabular}
}
\vspace{10pt}
\caption{Direction for included effects, selected models.  Positive coefficients are shown in green, negative in orange; effects not included are blank.  Most mechanisms show consistent effects across the networks in which they are present. \label{tab_effectincl}}
\end{table}

RRecSnd is a term for recency of receiving a communication from an alter affecting the actor’s current rate of sending to that particular alter, while RSndSnd is the recency of sending a communication to an alter affecting the actor’s rate of sending to that same alter. For the networks in which they are present, the mean value is 1.82 for RRecSnd and 2.63 for RSndSnd. Both effects are always positive when they appear in a final model. This would suggest that these cognitive mechanisms do govern some of the observed conversational patterns, with individuals relying on recall when taking an action to address individuals in the network.

FrPSndSnd is a general effect for persistence, which is the fraction of an actor’s past sending actions to a particular alter affecting their future rate of sending to that alter; this effect is always negative when present, with a maximum value of -0.50, a minimum value of -2.88, and a mean value of -1.36. This term being negative suggests an overall movement of conversation, rather than individuals continuously addressing only a small proportion of the network.

OTPSnd and ITPSnd are both triadic two-path effects, measuring the effect of the number of out-bound or incoming two-paths from actor to alter or alter to actor on the  rate of actors sending to that alter. OTPSnd ranges from -0.12 to 0.52, with a mean of 0.15, while ITPSnd ranges from -0.35 to 0.60, with a mean value of 0.03. This suggests that networks vary in their three-way communication styles, with some likely to observe triadic closure, while others trend away from these patterns. OSPSnd and ISPSnd are triadic terms for out-bound and incoming shared partner effects, where the number of shared partners between actor and alter affect the future rate of sending to that alter. OSPSnd ranges from -0.38 to 0.21, with a mean of -0.01. ISPSnd has a range between -0.30 and 0.22, with a mean at 0.01, which again suggests wide variation in triadic communication structure.

\subsection*{Hub-Forming Mechanisms}

Figure~\ref{fig_coef} provides a visual representation of the posterior modes and their asymptotic 95\% posterior intervals for the event model parameters for each of the 17 networks. Line segments that are darkly colored represent the specialist networks, while the lightly colored lines represent their non-specialist counterparts. Despite some small differences between specialists and non-specialists, in particular their slight difference in magnitude for preferential attachment, the effects are overwhelmingly similar between the two groups. Effects are especially similar in regards to direction, and in the majority of cases the magnitude of the effects are also strikingly similar. This can also be seen in  Table~\ref{tab_effectincl}, which clearly shows strong similarities in patterns across networks.  Indeed, figure~\ref{fig_coef} suggests that some effects may be even more consistent than Table~\ref{tab_effectincl} would indicate, since some apparently reverse-signed terms have relatively uncertain signs (as evidenced by 95\% posterior intervals that include 0).

Overall it appears that, when the respective mechanism is active in a given network, recency on receiving, recency on sending, and all of five conversational inertia terms (PSAB-BA, PSAB-BY, PSAB-XA, PSAB-XB, PSAB-AY) enhance event probability. Likewise, persistence always has a negative influence.  ICR and preferential attachment effects are nearly always positive (with the former having fairly similar size), but some networks show negative modal estimates (though the 95\% PIs are wide and cross zero, suggesting that the direction of effect is somewhat uncertain).  We observe that preferential attachment effects tend to be hard to estimate, with large posterior uncertainties, making quantitative comparisons across cases difficult. However, the overwhelming majority of preferential attachment effects are positive, with two networks as possible exceptions (though their 95\% posterior intervals include positive values).  The triadic structure terms (OTP, ITP, OSP, ISP) seem to have small magnitude effects of varying direction, and in some cases are of uncertain sign.  Their persistent inclusion and predominantly positive effects indicate systematic tendencies towards triadic closure, but it is possible that these reflect relatively general pressure for cohesive interaction that is not greatly sensitive to the type of triadic structure involved.

Considering these effects in more detail, we find that for the conversational inertia terms have the following positive effects on event probability: PSAB-BA, call and response, the most ubiquitous term found in every optimized final model, on average has an effect of 6.8, a minimum of 2.9 and a maximum effect of 9.9. PSAB-XA, source attraction, found in all but one model, has an effect mean of 3, a min of 0, and a max of 3.63. The PSAB-BY, ``handing off” term, is also found in all but one model, with a mean effect of 2.4, a min of 0, and a maximum value of 3.9. The fourth participation shift effect, found in 88\% of all models, is the PSAB-AY, sequential address term. Sequential address has a mean effect of 2.4, a min effect of 0, and a maximum value of 4. Finally, PSAB-XB, the effect of turn usurping, is found in 70\% of the models and accounts for a mean effect of 1.4, a min of 0, and a maximum of 2.9.

The effect of preferential attachment was found to be in 82\% of the optimized final models. The mean effect for preferential attachment, where present, has a mean effect of 3.5, with a maximum effect of 11.4.

The covariate effect for receiving and sending by ICRs appears in 65\% of all networks, with a minimum effect of -0.67, a mean effect of 0.52, and maximum effect of 1.7. Because of the way this term is specified, if two individuals of ICR status in the network communicate to one another the event probability effect is doubled: for instance, in the case of PATH police which has the greatest effect for ICR, the effect for two ICRs communicating would be $1.7\times2 = 3.4$.  Those networks with high-certainty coefficients have positive effects, but we see two cases with small negative posterior means and posterior intervals that place considerable mass in the positive direction.   

Finally, our analyses showed no systematic differences between the average coefficient of specialized and non-specialized networks. Therefore, at least in terms of effect magnitude and direction, specialized and non-specialized networks during a disaster event are more similar than they are different, despite what we might otherwise expect.

\subsection*{Knock-Out Experiment Results}

\begin{table}
\centering
\setlength\tabcolsep{0.2cm}
\resizebox{\columnwidth}{!}{%
\begin{tabular}{lrrrrr}
  \hline
 & Full & PA & PS & ICR & All \\
 & Model & Removed & Removed & Removed & Removed\\
  \hline
PATH Radio Comm & 0.68 &  & 0.23 &  & 0.23 \\ 
  Lincoln Tunnel Police & 0.37 & 0.40 & 0.35 & 0.14 & 0.14 \\ 
  Newark Command & 0.99 &  & 0.19 & 0.51 & 0.10 \\ 
  Newark Police & 0.25 & 0.29 & 0.09 &  & 0.11 \\ 
  Newark CPD & 1.25 & 0.98 & 0.41 & 0.49 & 0.07 \\ 
  Newark Operations Terminals & 1.72 & 0.27 & 0.05 &  & 0.04 \\ 
  Newark Maintenance & 0.55 & 0.44 & 0.20 & 0.35 & 0.08 \\ 
  PATH Control Desk & 0.80 & 0.74 & 0.16 & 0.25 & 0.06 \\ 
  NJSPEN 1 & 0.54 & 0.37 & 0.09 & 0.39 & 0.08 \\ 
  NJSPEN 2 & 1.11 & 0.58 & 0.26 &  & 0.14 \\ 
  WTC Operations & 0.64 & 0.57 & 0.08 &  & 0.08 \\ 
  WTC Police & 0.95 & 0.57 & 0.13 & 0.61 & 0.04 \\ 
  WTC Vertical Trans & 0.51 &  & 0.10 & 0.46 & 0.10 \\ 
  Newark Facility Management & 1.95 & 0.38 & 0.08 & 1.82 & 0.06 \\ 
  PATH Police & 1.86 & 1.66 & 0.26 & 0.20 & 0.04 \\ 
  WTC Security & 1.27 & 0.48 & 0.07 & 1.32 & 0.07 \\ 
  WTC Maintenance Electric & 1.97 & 0.77 & 0.16 &  & 0.11 \\ \hline
  Mean & 0.97 & 0.57 & 0.16 & 0.54 & 0.09 \\ 
   \hline
\end{tabular}%
}
\vspace{10pt}
\caption{Mean Theil index before and after mechanism knock-out. Full model includes all AICc-selected terms; for removed terms, PA=preferential attachment, PS=P-shifts, ICR=ICR covariate, all=all hub-forming mechanisms. \label{tab_theil} } 
\end{table}

\begin{table}
\centering
\setlength\tabcolsep{0.2cm}
\begin{tabular}{lrrr}
  \hline
 & \multicolumn{1}{c}{PA} & \multicolumn{1}{c}{P-Shifts} & \multicolumn{1}{c}{ICR}\\
  & \multicolumn{1}{c}{Removed} & \multicolumn{1}{c}{Removed} & \multicolumn{1}{c}{Removed} \\ \hline
  PATH Radio Comm &  & -65.37$^{***}$ &  \\ 
  Lincoln Tunnel Police &   7.47 &  -3.99 & -60.90$^{***}$  \\ 
  Newark Command &  & -80.96$^{***}$ & -48.71$^{***}$ \\ 
  Newark Police &  13.36 & -63.29$^{***}$ &  \\ 
  Newark CPD & -21.61$^{***}$ & -67.52$^{***}$ & -61.15$^{***}$ \\ 
  Newark Operations Terminals & -84.19$^{***}$ & -97.24$^{***}$ &   \\ 
  Newark Maintenance & -19.86$^{*}$ & -63.97$^{***}$ & -36.11$^{***}$ \\ 
  PATH Control Desk  &  -7.90$^{*}$ & -80.18$^{***}$ & -69.11$^{***}$  \\ 
  NJSPEN 1 & -31.81$^{*}$ & -82.69$^{***}$ & -28.66$^{*}$  \\ 
  NJSPEN 2 & -48.08$^{***}$ & -76.62$^{***}$ &   \\ 
  WTC Operations  & -10.23 & -86.91$^{***}$ &  \\ 
  WTC Police  & -39.94$^{***}$ & -86.25$^{***}$ & -36.30$^{***}$  \\ 
  WTC Vertical Trans  &  & -80.26$^{***}$ &  -9.71  \\ 
  Newark Facility Management  & -80.69$^{***}$ & -95.67$^{***}$ &  -6.81  \\ 
  PATH Police & -10.68$^{**}$ & -85.81$^{***}$ & -89.47$^{***}$ \\ 
  WTC Security & -62.06$^{***}$ & -94.14$^{***}$ &   4.12  \\ 
  WTC Maintenance Electric & -60.64$^{***}$ & -91.66$^{***}$ &  \\  \hline
  Mean  & -30.46 & -72.36 & -36.90 \\
   \hline
  \multicolumn{4}{c}{$^{*}$ $p<0.05$; $^{**}$ $p<0.01$; $^{***}$ $p<0.001$} \\
\end{tabular}
\vspace{10pt}
\caption{Percentage change in Theil index of communication volume mechanism knock-out. Full model includes all AICc-selected terms; for removed terms, PA=preferential attachment, PS=P-shifts, ICR=ICR covariate, all=all hub-forming mechanisms. $p$-values reflect two-sample $t$-tests (knock-out vs. full model). \label{tab_delta}}  
\end{table}

Our initial analysis suggests that the three hub-forming covariates of preferential attachment, ICR, and conversational inertia can all matter in terms of predicting the local conversational patterns, but does not directly address how these mechanisms contribute to hub formation (an emergent, macroscopic outcome of social microdynamics).  Here we employ simulation to investigate this latter question.

\begin{figure}
\centering
\includegraphics[width=\textwidth]{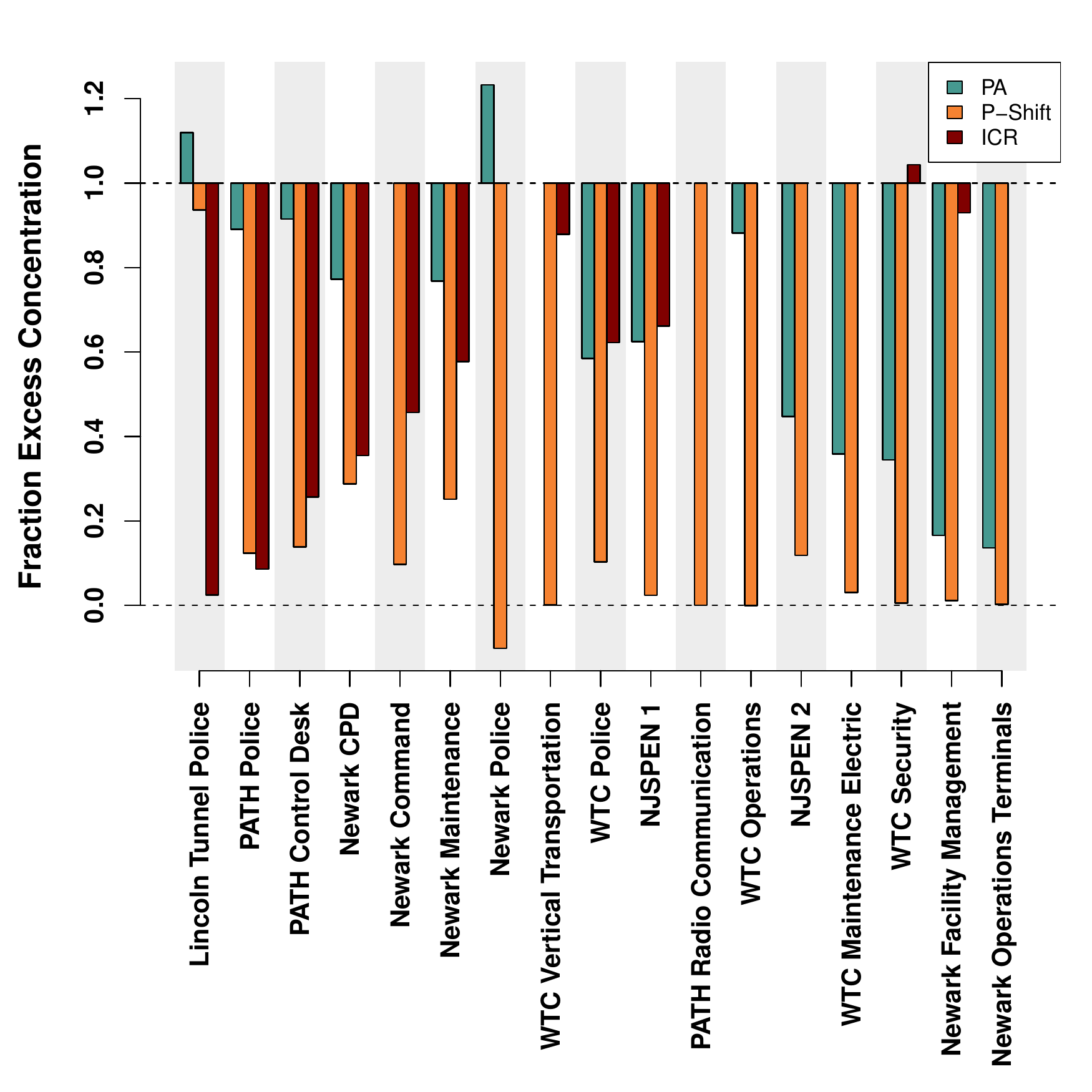}
\caption{Excess concentration in communication volume (above no-hub mechanism baseline) as a function of knock-out condition; values of 1.0 correspond to full model.  Networks differ in importance of ICR and PA effects for hub formation, while removing P-shift effects nearly always greatly reduces concentration. (Note: concentration outside the 0-1 interval is possible.) 
\label{theil_excon.pdf}}
\end{figure}

 Figure~\ref{theil_excon.pdf} provides an intuitive visual representation of the results from simulations of each network (the corresponding Theil index values can be found in Table~\ref{tab_theil}).  Here, we define the ``baseline'' concentration level when no hub effects are included as 0 (no ``excess'' concentration), normalizing all concentration values by affine transformation so that the level of concentration when all hub effects are included is set equal to 1 (100\% of the ``excess'' concentration produced by all effects in tandem).  Given this scale, we can now compare across networks the relative amount of excess concentration that remains when we suppress each class of hub effect (vertical bars).  Note that while the fraction of concentration observed after knocking out an effect is usually in the [0,1] interval, it does not have to be: values higher than the ``all hub effects'' level, or lower than the ``no hub effects'' level, can occur when either (1) a nominally hub-promoting mechanism actually inhibits hub formation in a specific network, or (2) nonlinear interactions between dynamic mechanisms lead to nonmonotonic outcomes. The most obvious cases here involve the two networks (Lincoln Tunnel Police and Newark Police) with negative preferential attachment effects; since the PA mechanism is on average hub \emph{suppressive} for these two cases, knocking it out actually enhances hub formation.  For convenience in interpretation, networks in Figure~\ref{theil_excon.pdf} are ordered using a 1-dimensional non-metric multidimensional scaling of the Euclidean distance between their respective concentration levels; this highlights both similarities and differences across networks.  Percentage change in values following knock-out can be found in Table~\ref{tab_delta}.
   
Overall, conversational inertia terms account for the largest individual factor reducing the Theil index when they are systematically removed from the models, with these terms comprising the strongest hub-forming mechanism in 88\% of the networks, and knocking them out reduces hub formation in all of the networks we analyzed. The mean reduction for participation shifts removal was an 88\% reduction, with a median reduction of 97\%. For one network, Newark Police, we find that by removing the conversational inertia p-shifts terms while keeping preferential attachment, that the network has a 110\% reduction in the theil index -- thus we are seeing a 10\% greater reduction for just removing p-shifts than a model with none of the hub-forming terms.  (This appears to be driven, as noted, by the presence of a hub-suppressive PA effect.)

Knocking out ICR effects in these networks tends also to reliably reduce hub-formation, with this effect seen in 91\% of the networks the effect appears in. Across all networks, where the ICR term was eligible to be removed, we found a mean reduction of 30\% in the Theil index. For PATH Police and Lincoln Tunnel Police, we find that ICR removal provides a larger reduction in hub-formation than conversational inertia term’s effect, at 91\% and 97\% respectively. We also find that in the WTC Maintenance Electric network removing ICR has a slight 4\% increase in the Theil index for inequality of total communication volume.

Preferential attachment (PA) follows suit with conversational inertia and ICR effects, where 85\% of the models where preferential attachment was removed caused a reduction in the Theil index of inequality of total communication volume. The average effect of removing PA, in the networks it is present in, is a 28\% reduction in the Theil index for communication volume, with the largest reduction in the Newark Operations Terminals at 86\%. However, as noted above, for Lincoln Tunnel Police and Newark Police, we find that by removing the PA term the resulting models have increases in the Theil index, of 12\% and 23\%. Therefore, it appears that on average conversational inertia terms (participation shifts) have the largest impact on system level hub formation (88\%), followed by ICRs (30\%), and preferential attachment (23\%). 

Looking across the effects in Figure~\ref{theil_excon.pdf}, we observe that the impact of p-shifts is consistent, but that there is a general trend in which ICR effects trade off against PA effects: networks that see large reductions in hub formation when ICR effects are knocked out show smaller effects for knocking out PA effects, and vice versa.  This suggests a mechanism for the contrast between ICRs and emergent coordinators discussed by \citet{petrescu-prahova.butts:ijmed:2008}, with preferential attachment arising as a heuristic for filling coordinator roles when ICRs are unable to carry the requisite load. It should be noted in this regard that for many of these networks, positive effects for both ICR interaction and preferential attachment are present; clearly, however, the mechanisms are not equally critical to hub formation in all networks. 


\begin{table}
\centering
\setlength\tabcolsep{0.2cm}
\begin{tabular}{rllll}
  \hline
 & PA & PS & ICR & All \\
 & Removed & Removed & Removed & Removed\\
  \hline
Specialist & -24.17 & -71.25 & -45.87 & -84.97 \\ 
  Non-Specialist & -43.92 & -82.66 & -30.43 & -87.47 \\ 
  Small & -21.14 & -72.69 & -55.76 & -83.44 \\ 
  Medium & -47.08 & -88.39 & -24.42 & -90.67 \\ 
  Large & -35.44 & -70.35 & -36.63 & -85.42 \\ 
   \hline
\end{tabular}
\vspace{10pt}
\caption{Mean percentage change in Theil index by group, under mechanism knock-out.  For removed terms, PA=preferential attachment, PS=P-shifts, ICR=ICR covariate, all=all hub-forming mechanisms. \label{tab_spectheil}} 
\end{table}

Though we do see some patterns of difference, they do not necessarily fall out along lines that are \emph{a priori} obvious.  In order to determine whether there were differences in system level hub formation between our specialized and non-specialized networks, we compared the mean reduction in Theil indices for each knock-out simulation across the two groups, in Table~\ref{tab_spectheil}. In order to determine whether these differences were significant between groups, we performed a two-sample $t$-test. We find that none of the changes in the Theil indices between specialized and non-specialized networks within our simulation results were significantly different, suggesting that each mechanism has the same impact, on average, in both categories.

We also compare the differences in the mean reduction in Theil indices for each knock-out simulation between differently sized networks, in Table~\ref{tab_spectheil}. 
To test whether these size differences were significantly different we conducted a 
Kruskal-Wallis one-way analysis of variance for three differently sized groups. As with the differences among specialist and non-specialist categories, we find no significant differences between the size of the networks and the mean reduction in Theil indices. 

\section*{Discussion}

In this study we focused on understanding the process of emergent coordination within the context of an unfolding disaster. We did so by investigating networks from the 2001 World Trade Center disaster, providing empirical evidence for the role of conversational norms, preferential attachment, and institutional status in shaping both the local and system level structure and dynamics. From prior studies, it was known that the WTC radio networks had hub-like structures that emerged during the communication process. It was also known that while an individual in an institutionalized coordinator role would have a greater likelihood of inhabiting a hub position, it was more often the case that these hubs were emergent, meaning they were occupied by individuals without institutional coordinator status. Our analysis takes this prior work one step further by modeling and simulating these networks, allowing us to understand which factors tend to guide communication, and how these same factors affect the overall structure. We did this by determining which mechanisms from the space of plausible effects were active in each network, and by estimating the strength and direction of effects that were present.  We then conducted a knock-out simulation experiment to understand how the three classes of potential hub formation mechanisms impact emergent structure, comparing network Theil indices of communication volume with and without each of the respective effects. Lastly, we compared our results from both analyses across various groups to search for systematic differences by specialization and size. Overall, we find that despite the differing contexts, environments, roles, specializations, and even sizes, the results across all networks are far more similar than different.  When mechanisms appear, they generally have the same sign (and usually are of comparable magnitude).  Deviations from these central tendencies seem idiosyncratic, and do not suggest a clear pattern of differences by size or specialization.
In terms of local effects on conversation, there is a dominance of p-shift effects being present in all networks, while ICR and preferential attachment only appear in particular cases; this showcases the significance of radio SOP and conversational norms. Again, it is striking that regardless of whether individuals are communicating at ground zero, or responding to an event happening in the next state over, the communication patterns we observe are surprisingly robust.

In terms of the simulation knock-out experiment results, the first general observation is that almost all of the findings are significant, meaning that when you systematically take out any of the hypothesized hub-forming mechanisms, networks on average show large reductions in the Theil index of communication volume from their baseline state. Further, we find that the largest influencing factor is conversational inertia, which represents an 88\% decrease on average in the Theil index of communication volume when the effect is knocked out. This was followed by the ICR effect and the preferential attachment effect, with reductions of 30\%, and 28\% respectively. In particular, these findings suggest that conversational norms are strong drivers of emergent coordination \citep[as suggested by ][]{gibson.et.al:po:2019}, even during an unfolding disaster -- findings that also accord with participation shift usage of pilots during the unfolding Air France 447 incident \citep{david.schraagen:ctw:2018}. While both ICR and preferential attachment do seem to affect the formation of coordination hubs, they do not seem to be the dominant force. The low impact from our ICR term tells us that individuals are likely to form hubs and coordinate organically in situations of disaster and that ICRs are not a necessary requirement for emergency coordination to take place. Preferential attachment's lower level of impact on our index tells us that while preferential attachment has an important role to play in disaster communication, it is not the sole driver in every context. Our results lead us to believe that more consideration should be given to alternative forms of coordination in future literature.

By comparing REM coefficients as well as simulated behavior, we show that coordination arises in these communication networks through surprisingly similar means; what we are seeing is an emergent macro sociological phenomenon of human communication during a hazard event. It seems that by and large the most important factor driving the emergent coordination is conversational inertia. In particular, the AB-BA call and response participation shift was found in every single network after model optimization, while the AB-XA source attraction, and AB-BY “handing off” terms were present in all but one of the networks - compare that to the preferential attachment term appearing in 14 networks, and the ICR term appearing in only 11. Because of the robustness of our findings across various categorizations, we believe these results would likely reappear in other hazard or disaster contexts. Further, we believe that this might even transfer in part to non-disaster contexts that are particularly influenced by conversational norms. Researchers going back to the late 1960’s and 1970’s have documented the importance of conversational norms in governing social interactions \citep{schegloff:aa:1968,schank:cs:1977} ; these results show that these humble aspects of micro-interaction have important macroscopic consequences, and that those hold even in the midst of an unfolding disaster.

\section*{Future Studies}

An obvious question for future work is whether the patterns seen here generalize to other hazard communication contexts. While these results appear to be quite robust - with regular patterns across networks of different size, composition, and task orientation - all networks reflect groups responding at the same time to the same event using the same technology, and some aspects may differ in other settings. For instance, in our study’s context, individuals were restricted to a single channel of voice communication, greatly amplifying the importance of conversational norms for communication.  It is possible that other media with less dependence upon such norms might yield a correspondingly reduced impact of conversational norms on hub formation, or simply less hub formation overall. In this vein, it would be natural to probe the emergence of coordination in other single channel communication media, like singular chat rooms (NWSChat), and in multi-channel communication media, like slack or discord. Do these newer forms of communication media create new kinds of norms, or hinder certain kinds of interactions?

In our study we found that preferential attachment, despite being a popularly studied mechanism for social network hub-formation, was not the dominant contributing factor for these communication networks. This leads us to believe that in future studies, where hub-formation in networks is of interest, more careful theoretical consideration should be paid to alternative mechanisms. In this case, conversational inertia played the largest role in coordination across multiple communication contexts, a novel finding for disaster communication networks.  It is plausible that concentration in other networks may also stem from similar microdynamics.  Since such behavior is ``invisible'' in conventional network studies, caution is needed when interpreting macroscopic patterns; formal development that links unobserved micro-processes with network structure (e.g., \citet{butts:jms:2020a} in an ERGM setting, or \citet{snijders:sm:2001} for panel dynamics) is helpful in this regard, but gaining data on microdynamics when possible is certainly advisable.

Finally, researchers may want to more broadly consider the implications of ``how'' an organization communicates, and what norms are set by and exist in certain media channels, as they could greatly shape the potential for organizations to deal with ongoing and future threats. This may involve taking an inventory of the various communication media an organization uses officially or unofficially, assessing how robust they are to infrastructure failure, and assessing whether secondary forms of communication may serve as primary forms in an active hazard or disaster event. Lastly, it should be considered how the alternative or secondary communication forms might hinder the effectiveness of their potential communication process.

\section*{Conclusion}

Even at this temporal remove, the WTC case remains unique in the lens it provides on the detailed dynamics of the emergency phase of a disaster, and it continues to offer lessons in the drivers of coordinative behavior.  In this special issue recognizing progress in REMs since their introduction, it seems apposite to provide a complete analysis of the data that motivated the initial work.  The accompanying data release will likewise be a useful resource for others who wish to investigate this important historical case. Our findings point to an emergent macro sociological phenomenon, finding that even in chaotic disaster events, and diverse contexts, actors in these networks by and large utilized the conversational norms we use in everyday life.

\label{lastpage}

\bibliography{WTCREM}

\end{document}